\documentclass[sigconf]{acmart}
\usepackage{multirow}
\usepackage{xcolor,colortbl}
\usepackage{graphicx}
\usepackage{framed}
\usepackage{balance}
\usepackage{comment}

\AtBeginDocument{%
  }

\copyrightyear{2026}
\acmYear{2026}
\setcopyright{cc}
\setcctype{by}
\acmConference[MSR '26]{23rd International Conference on Mining Software Repositories}{April 13--14, 2026}{Rio de Janeiro, Brazil}
\acmBooktitle{23rd International Conference on Mining Software Repositories (MSR '26), April 13--14, 2026, Rio de Janeiro, Brazil}
\acmPrice{}
\acmDOI{10.1145/3793302.3793567}
\acmISBN{979-8-4007-2474-9/2026/04}

\newcommand{\RqOne}{\textbf{RQ1:} \emph{To what extent can statistical regression models determine whether a PR created by a human or an AI agent will be merged?}}
\newcommand{\RqTwo}{\textbf{RQ2:} \emph{How do the features contributing to merge probability differ between PRs created by humans and those created by AI agents?}}
\newcommand{\RqThree}{\textbf{RQ3:} \emph{How do the features contributing to merge probability differ across PRs generated by different AI agents?}}

\begin{document}

\title{Let's Make Every Pull Request Meaningful: An Empirical Analysis of Developer and Agentic Pull Requests}


\author{Haruhiko Yoshioka}
\orcid{0009-0004-2353-1397}
\author{Takahiro Monno}
\orcid{0009-0001-4669-8411}
\affiliation{%
  \institution{Nara Institute of Science and Technology}
  \city{Ikoma}
  \state{Nara}
  \country{Japan}
}
\email{yoshioka.haruhiko.yi4@naist.ac.jp}
\email{monno.takahiro.mv3@naist.ac.jp}

\author{Haruka Tokumasu}
\orcid{0009-0006-5463-6305}
\author{Taiki Wakamatsu}
\orcid{0009-0000-4237-106X}
\affiliation{%
  \institution{Kyushu University}
  \city{Fukuoka}
  \state{Fukuoka}
  \country{Japan}
}
\email{tokumasu@posl.ait.kyushu-u.ac.jp}
\email{wakamatsu.taiki.448@s.kyushu-u.ac.jp}

\author{Yuki Ota}
\orcid{0009-0001-3657-161X}
\affiliation{%
  \institution{Ritsumeikan University}
  \city{Ibaraki}
  \state{Osaka}
  \country{Japan}
}
\email{is0607ih@ed.ritsumei.ac.jp}

\author{Nimmi Weeraddana}
\orcid{0009-0002-6550-4181}
\affiliation{%
  \institution{University of Calgary}
  \city{Calgary}
  \state{Alberta}
  \country{Canada}
}
\email{nimmi.weeraddana@ucalgary.ca}

\author{Kenichi Matsumoto}
\orcid{0000-0002-7418-9323}
\affiliation{%
  \institution{Nara Institute of Science and Technology}
  \city{Ikoma}
  \state{Nara}
  \country{Japan}
}
\email{matumoto@is.naist.jp}

\renewcommand{\shortauthors}{Yoshioka et al.}

\begin{abstract}
  The automatic generation of pull requests (PRs) using AI agents has become increasingly common.
  Although AI-generated PRs are fast and easy to create, their merge rates have been reported to be lower than those created by humans.
  In this study, we conduct a large-scale empirical analysis of 40,214 PRs collected from the \textsc{AIDev} dataset.
  We extract 64 features across six families and fit statistical regression models to compare PR merge outcomes for human and agentic PRs, as well as across three AI agents.
  Our results show that submitter attributes dominate merge outcomes for both groups, while review-related features exhibit contrasting effects between human and agentic PRs.
  The findings of this study provide insights into improving PR quality through human–AI collaboration.
\end{abstract}




\maketitle

\section{Introduction}

In software development, creating pull requests (PRs) requires substantial effort from developers~\cite{Gousios2016}. Beyond implementing code changes, developers must also provide contextual information, such as clear statements of intent, references to related issues, and responses to peer-reviewer feedback.
Existing literature~\cite{Gousios2014,Tsay2014,Zhang2023,Iyer2021,Meijer2025,Alami2022} has extensively analyzed the factors influencing the acceptance or rejection of developer-authored PRs.
For example, Gousios et al.~\cite{Gousios2014} have revealed that PR acceptance is influenced by change size, file activity, project scale, contributor's track record, and the review process.
Furthermore, others have shown that the richness of descriptions, the nature of discussions, and the presence of tests~\cite{Tsay2014} impact the acceptance of developers' PRs.

To help developers create PRs, Artificial Intelligence~(AI) agents are used to automatically create change sets and submit PRs~\cite{Li2025aiteammates}.
However, AI-generated PRs (i.e., \textit{Agentic PRs}) are not without their limitations~\cite{Cotroneo2025,Bukhari2023,Chouchen2024,Cihan2025,Li2025aiteammates}.
In fact, AI-generated code has been reported to be structurally simple and repetitive, prone to containing vulnerabilities~\cite{Cotroneo2025}, and identifiable by specific lexical and syntactic features~\cite{Bukhari2023}.
Besides, PRs that include at least one ChatGPT shared link take longer to be closed (merged or abandoned)~\cite{Chouchen2024}. 
Then, Li et al.~\cite{Li2025aiteammates} demonstrated that the acceptance rate of Agentic PRs
(i.e., the proportion of Agentic-PRs that are successfully merged) in \textsc{AIDev} dataset
is significantly lower than that of humans. 
However, little is known about the quality and acceptability of such PRs compared to those authored by humans~\cite{Li2025aiteammates, Cotroneo2025}.


The primary goal of this study is to enhance PR quality by leveraging the complementary strengths of humans and GenAI.
To compare the factors that characterize the accepted PRs that were created by humans and agents, we first extract 64 features across 6 families from a dataset of 6,618 human-authored PRs and 33,596 agentic PRs from the \textsc{AIDev} dataset~\cite{Li2025aiteammates}. 
These features include change set size, number of files added/deleted/changed, task types, external links, and AI agent type (e.g., OpenAI Codex, Copilot).
For humans and each of the AI models in \textsc{AIDev} dataset, we fit and analyze regression models to uncover the features of PRs that make them \emph{difficult for GenAI agents to get their PR merged but easy for humans} (and vice versa).
By gaining insights into such features, we provide actionable evidence regarding the behavior of AI agents in real-world projects, offering guidelines for developers and project managers to optimize development processes where AI and humans collaborate~\cite{Li2025aiteammates}.
This study addresses the following three research questions (RQs):
\vspace{0.25em}

\noindent
\textbf{\RqOne}
As a foundation for our comparative analysis between humans and AI agents, we analyze the extent to which the constructed regression models can perform on real-world PRs.
\vspace{0.25em}

\noindent
\textbf{\RqTwo}
This RQ aims to uncover the respective strengths and weaknesses of humans and agents in getting their PRs merged. We investigate the explanatory power of each feature family in our statistical models to identify which ones contribute most to PR merging, providing insights into which aspects are most critical for humans versus agents.
\vspace{0.25em}

\noindent
\textbf{\RqThree}
In this RQ, we compare the importance of feature families in predicting the merge outcomes of agent-authored PRs across different AI agents, with the goal of providing insights into which AI agents are better suited for specific development contexts.

\vspace{0.25em}
\noindent
Our key findings reveal that: (1) submitter attributes dominate merge outcomes for both human and agentic PRs; (2) increased review activity is associated with higher merge likelihood for human PRs but lower for agentic PRs; and (3) AI agents show distinct merge patterns.



\begin{figure*}[t]
\centering
\includegraphics[width=1.0\textwidth]{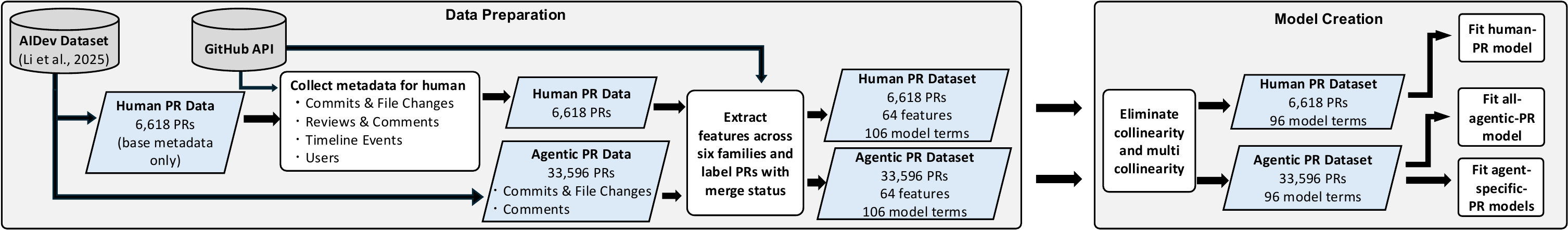}
\caption{An overview of our study design}
\label{fig:data_collection}
\end{figure*}

%
%
%
%
%
\section{Study Design}

\begin{table}[t]
  \centering
  \caption{Families of Features}
  \label{tab:features}
  \small
    \resizebox{0.7\columnwidth}{!}{
  \begin{tabular}{lr}
  \hline
  \rowcolor[HTML]{DDDDDD}
  \textbf{Family} & \textbf{\# Features} \\
  \hline
  PR Change Size \& Commit Characteristics & 18 \\
  PR Description & 8 \\
  Submitter Attributes & 6 \\
  Repository Attributes & 3 \\
  Issue Linkage \& Surrounding Context & 3 \\
  Reviews \& Discussion & 26 \\
  \hline
  \textbf{Total number of features} & \textbf{64} \\
  \hline
  \end{tabular}
  }
\end{table}

In this section, we describe our method for data preparation (Section~\ref{sec:dp}) and the model creation (Section~\ref{sec:mc}). Figure~\ref{fig:data_collection} overviews our study design.
\vspace{0.25em}

\noindent
\subsection{Data Preparation}\label{sec:dp}

\textbf{Collect metadata from \textsc{AIDev} dataset.} We begin with the \textsc{AIDev} dataset (Nov.\ 2025) released by Li et al.~\cite{Li2025aiteammates}, which consists of 932,791 Agentic PRs created by five major autonomous AI agents.
Of these, 33,596 include rich metadata such as comments, issues, reviews, commits, repositories, timelines, and user information.

\textbf{Collect metadata for human-authored PRs.} For comparison purposes, the \textsc{AIDev} dataset contains 6,618 human-authored PRs created by 2,515 developers (hereafter referred to as \textit{human PRs}) across 818 repositories. However, for these human PRs, the \textsc{AIDev} dataset provides only repository-level information and task type, and does not include other metadata such as comments, reviews, or commits.
To enable a fair comparison between agentic PRs and human PRs, it is necessary to use the same set of features for both. Therefore, we collect the same set of metadata for human PRs as those available for agentic PRs by using the same GitHub API endpoints as Li et al. when constructing the \textsc{AIDev} dataset~\cite{Li2025aiteammates}.

\textbf{Extract features from human PRs and agentic PRs, covering distinct families of features.} We use our dataset of metadata to extract 64 features across six families (while consulting prior work~\cite{Zhang2023,soares2015,Tsay2014}) to represent different dimensions of a PR, ranging from code changes to social interactions.
Table~\ref{tab:features} shows these families of features.
For example, the family that covers \textit{PR Change Size \& Commit Characteristics}~\cite{Zhang2023} includes 18 features, such as \textit{commit count, unique committers}, and \textit{average changes per commit}. Similarly, the family that covers \textit{PR Description} contains eight features, including \textit{message token count, message character count}, and \textit{average line count of messages}.
Then, the family of  Submitter Attributes~\cite{Zhang2023} contains features, such as the \textit{number of prior reviews made by the PR submitter, social distance} (i.e., whether the submitter follows the user who closes the pull request), and the \textit{submitter's follower count}. 
The full list of 64 features is available in our online appendix.\footnote{\label{fn:replication}\url{https://zenodo.org/records/18373332}}
After encoding all the categorical features with one-hot encoding, our dataset expanded to 106 model terms.
\vspace{0.25em}

\textbf{Label PRs with merge status.}
To be able to analyze differences in families of features that contribute to merge status, we determine the merge status of each PR by checking if the \texttt{merged\_at} field of a PR is \texttt{NULL}~\cite{Li2025aiteammates}. If so, the PR is considered not merged; otherwise, it is regarded as merged.
Note that this labelling process is applied to both agentic PRs and human PRs.

\subsection{Model Creation}
\label{sec:mc}

Our goal is to study what characterizes merged PRs created by humans and agents. 
Therefore, we select a statistical regression modelling approach which, unlike many other classification techniques, emphasizes interpretability. Such models (e.g., logistic regression) provide clear insights into how different factors influence outcomes, making them suited for nuanced analysis. Below, we discuss how we create these models. 

\textbf{Eliminate collinearity and multicollinearity.}
Collinear terms distort importance estimates~\cite{hastie2009elements}.
We use Variance Inflation Factor (VIF)~\cite{montgomery2021introduction} to detect multicollinearity in our regression models, as VIF captures higher-order dependencies that pairwise correlation coefficients may fail to identify.
Applying the commonly used threshold of VIF~$<10$~\cite{Obrien2007}, we eliminate ten model terms (two due to perfect collinearity and eight with VIF~$>10$), reducing the number of model terms across the feature families from 106 to 96.



\textbf{Fit the models.}
We fit logistic regression models using the set of 96 model terms that remained after the collinearity elimination step.
First, we train two main models using the full datasets: one on all human PRs and one on all agentic PRs.
Note that our dataset contains Agentic PRs created by five AI agents: OpenAI Codex ($N=21,799$), GitHub Copilot ($N=4,970$), Cursor ($N=1,541$), Devin ($N=4,827$), and Claude Code ($N=459$). 
Therefore, we further fit five additional models by partitioning the Agentic PR dataset by agent name and fitting a separate logistic regression model for each agent.
Our dataset and replication package is available online.\textsuperscript{\ref{fn:replication}}

To answer RQ1, we analyze model fitness, which is a prerequisite to ensure that our models provide reliable insights. We then compare the explanatory power across different feature families in the human-PR model and all-agentic model to answer RQ2, and examine agent-specific PR models to address RQ3.

\section{Study Results}
In this section, we present the approach and results for our RQs.

\begin{table}[t]
  \centering
  \caption{Prediction Performance}
  \label{tab:performance}
  \small
  \resizebox{\columnwidth}{!}{
    \begin{tabular}{lccccccc}
      \hline
      \rowcolor[HTML]{DDDDDD}
      Metric    & Human     & All-agentic & OpenAI    & Copilot   & Cursor    & Devin     & Claude    \\
      \rowcolor[HTML]{DDDDDD}
               &        &          & Codex    &           &           &           & Code      \\
      \hline
      AUC      & 0.878     & 0.960       & 0.997     & 0.775     & 0.969     & 0.810     & 0.904     \\
      Precision & 0.859    & 0.939       & 0.994     & 0.690     & 0.958     & 0.744     & 0.963     \\
      Recall   & 0.911     & 0.909       & 0.984     & 0.582     & 0.904     & 0.809     & 0.768     \\
      F1       & 0.884     & 0.923       & 0.989     & 0.631     & 0.930     & 0.775     & 0.850     \\
      Brier Score    & 0.119     & 0.073       & 0.014     & 0.193     & 0.063     & 0.176     & 0.113     \\
      \hline
    \end{tabular}
  }
\end{table}

\begin{table}[t]
  \centering
  \caption{Feature Family Importance: Human vs Agentic PRs}
  \label{tab:family_importance}
  \small
  \resizebox{\columnwidth}{!}{
  \begin{tabular}{l|rrr|rrr}
  \hline
  \rowcolor[HTML]{DDDDDD}
  & \multicolumn{3}{c|}{\textbf{Human PRs}} & \multicolumn{3}{c}{\textbf{Agentic PRs}} \\
  \rowcolor[HTML]{DDDDDD}
  \textbf{Family} & $\chi^2$ & \textit{df} & \textit{p} & $\chi^2$ & \textit{df} & \textit{p} \\
  \hline
  Submitter Attributes & 1,969.19 & 6 & * & 19,440.85 & 6 & * \\
  Reviews \& Discussion & 202.03 & 22 & * & 1,195.76 & 22 & * \\
  PR Change Size \& Commit & 118.55 & 17 & * & 1,078.55 & 17 & * \\
  Repository Attributes & 189.54 & 37 & * & 633.78 & 40 & * \\
  Issue Linkage \& Context & 4.00 & 3 & $\circ$ & 250.65 & 3 & * \\
  PR Description & 17.99 & 4 & * & 93.21 & 4 & * \\
  \hline
  \textbf{Entire model} & 2,853.86 & 96 & * & 25,528.61 & 96 & * \\
  \hline
  \multicolumn{7}{l}{\small *$p<0.001$ \quad $\circ$Not Significant ($p=0.262$)} \\
  \end{tabular}
  }
\end{table}

\subsection{\RqOne}
\label{sec:rq1}

\textbf{Approach.}
We evaluate the discriminatory power of our models using AUC~(0.5 = random, 1 = perfect), precision, recall, and F1 score (0 = worst, 1 = best), and calibration using the Brier score (0~=~best, 1 = worst), via five-fold cross-validation.



\vspace{0.25em}
\noindent\textbf{Results.}
Table~\ref{tab:performance} shows the performance of our logistic regression models. 
Accordingly, all the models surpass the baseline performance of a random guesser in terms of AUC when predicting whether a PR will be merged across all cases.
For example, the model for human PRs shows an AUC of 0.878, while the All-Agentic model achieves a higher AUC of 0.960.
Besides, the precision, recall, and F1-scores of our models are also closer to 1, demonstrating robust performance under class imbalance. 
In terms of the Brier score, all models achieved values close to zero, indicating well-calibrated probability estimates in their predicted merge probabilities. 
\textit{\textbf{Overall, our models are strong, suggesting that the observations derived from these models are meaningful and suitable for comparative analyses in upcoming RQs: RQ2 \& RQ3.}}


\subsection{\RqTwo}
\label{sec:rq2}
\textbf{Approach.}
In this RQ, we identify differences in strengths between humans and AI agents in contributing to PR merges. To this end, we use the human-PR model and the all-agentic-PR model to analyze which feature families are most important for predicting merge probabilities for human PRs and agentic PRs.

\begin{figure}[t]
    \centering
    \includegraphics[width=0.4\linewidth]{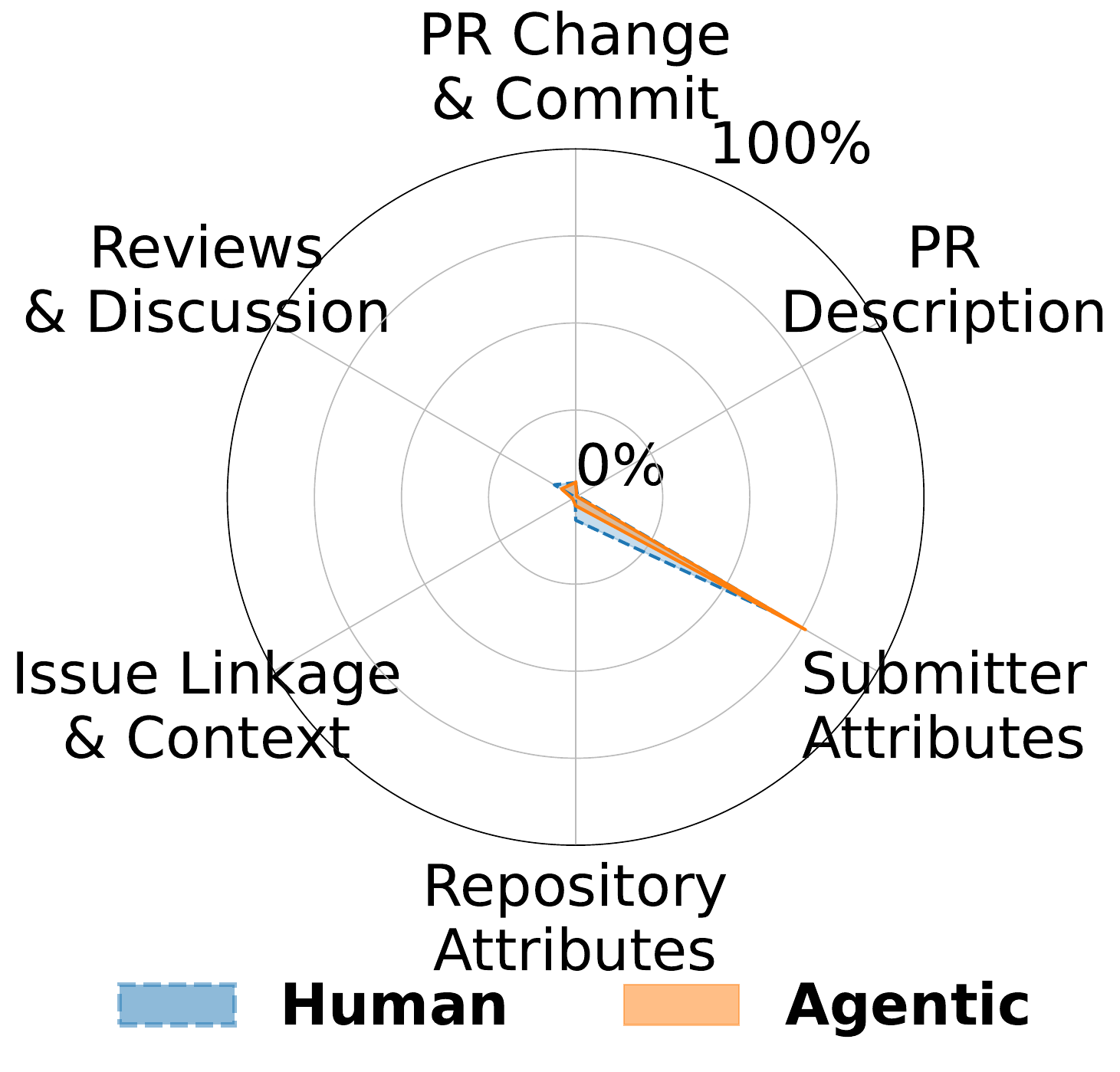}
    
    \caption{Feature family importance: human vs agentic PRs}
    \label{fig:rq2_radar}
\end{figure}

We use Likelihood Ratio (LR) $\chi^2$ chunk tests~\cite{harrell2015regression} to estimate feature family importance.
A larger LR $\chi^2$ value indicates greater loss of explanatory power when a family is removed, signifying higher importance in explaining merge outcomes.
We compare relative importance ($\frac{\chi^2\text{value of the family}}{\chi^2\text{value of the entire model}} \times 100\%$) between human and agentic PRs using radar charts.


\vspace{0.25em}
\noindent\textbf{Results.}
Table~\ref{tab:family_importance} compares the importance of feature families between human PRs and agentic PRs.
The $\chi^2$ column reports the LR $\chi^2$ statistic from chunk tests, ranking feature families by explanatory power. The df column indicates model terms per feature family after collinearity elimination. 
For comparison purposes, Figure~\ref{fig:rq2_radar} visualizes the relative importance of each feature family as a percentage of the total LR $\chi^2$ in the human-PR model and all-agentic-PR model.

Table~\ref{tab:family_importance} and Figure~\ref{fig:rq2_radar} show that submitter attributes dominate merge outcomes for both human PRs and agentic PRs, with relative importance of  $\frac{1,969.19}{2,853.86}$$\times$100=69.00\% and $\frac{19,440.85}{25,528.61}$$\times$100=76.15\%, respectively.
%
%
%
%
A feature-level analysis of this family reveals that when the contributor is also the integrator of a PR (i.e., when the value of \texttt{same\_user} feature is one), the odds of merging a PR increase by approximately 219 times for human PRs and by 5,419 times for agentic PRs, compared to cases where the contributor and integrator are different individuals. 
We find that 57.63\% of merged human PRs have \texttt{same\_user}=1, compared to 77.51\% of merged agentic PRs. This difference is also statistically significant (two-proportion z-test: $z = -29.38$, $p < 0.001$), confirming that agentic PRs are substantially more likely than human PRs to be merged by the submitter of that PR.

\begin{framed}
\noindent\textbf{Implication 1.} This finding provides insights into a potential risk: allowing the same entity (especially an AI agent) to both author and merge PRs may limit independent oversight.
Project maintainers may investigate such cases further and enforce third-party approval for PRs when deploying AI agents in development workflows.
\end{framed}

For both human PRs and agentic PRs, reviews and discussion attributes form the second most important feature family, with a relative importance of $\frac{202.03}{2,853.86}$$\times$100=7.08\% and $\frac{1,195.76}{25,528.61}$$\times$100=4.68\%, respectively. 
However, feature-level analysis shows that these signals carry different meanings across the two PR types.
%
%
For instance, each additional reviewer comment increases merge odds by 2.7\% for human PRs, but decreases merge odds by 2.8\% for agentic PRs.
%
%
%
In fact, 3.7\% of merged agentic PRs have more than three reviewers, whereas a higher percentage (i.e., 5.8\%) of unmerged agentic PRs have more than three reviewers.

\begin{framed}
\noindent\textbf{Implication 2.}
Our finding provides insights into another potential issue: extensive review activity around agentic PRs may signal the amount of correction required before merging. We encourage project maintainers to break down code changes into smaller, more manageable tasks for agents, enabling easier integration and code review.
\end{framed}

\subsection{\RqThree}
\label{sec:rq3}
\textbf{Approach.}
In this RQ, we aim to understand the characteristics of specific AI agents to provide insights into selecting the most suitable agent. 
We apply LR $\chi^2$ chunk tests to agent-specific models (Section~\ref{sec:mc}) to assess feature family importance.
Note that we exclude the logistic regression models for Cursor and Claude Code  because they exceed the degrees-of-freedom budget defined as  $\frac{\text{\# PRs in minority class}}{15}$~\cite{harrell1984regression,harrell1985regression}.
We then visualize $\chi^2$ statistics using radar charts,\textsuperscript{\ref{fn:replication}}
and identify the feature families that have the highest explanatory power in the remaining three models.

\vspace{0.25em}

\begin{figure}[t]
    \centering
    \includegraphics[width=1.00\columnwidth]{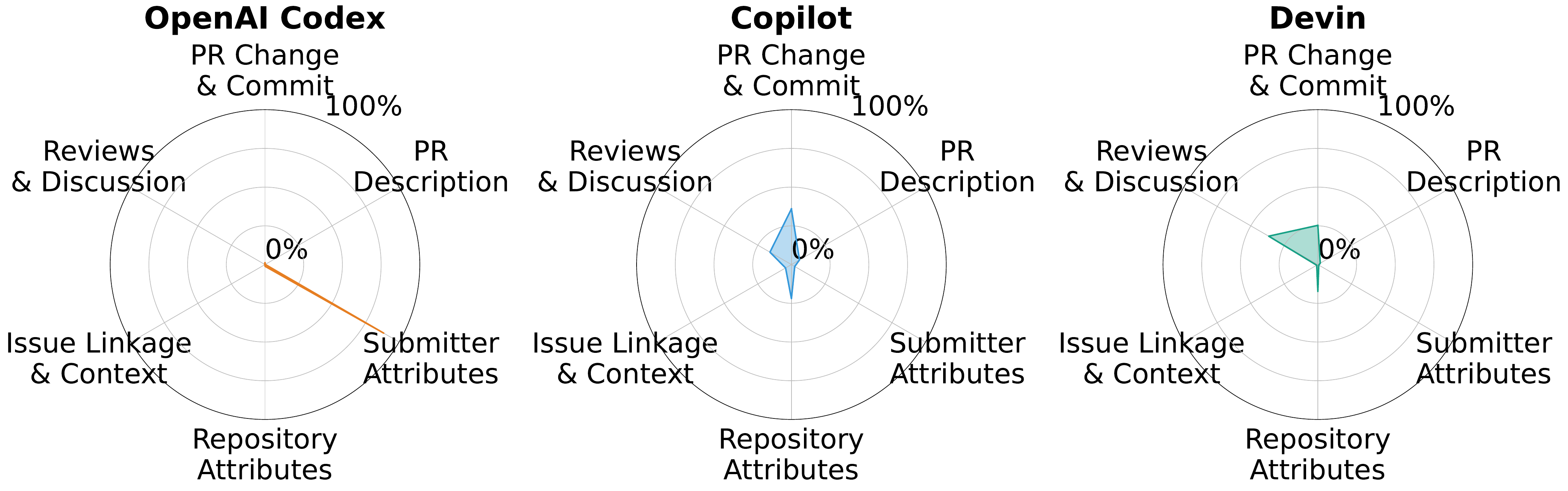}
    \caption{Feature family importance across AI agents}
    \label{fig:rq3_radar}
\end{figure}

\noindent\textbf{Results.}
Figure~\ref{fig:rq3_radar}, the radar charts, shows that different agents' PR merge likelihood is correlated with different feature families. 
In fact, OpenAI Codex’s most important feature family is submitter features, consistent with the RQ2 models, whereas Copilot is primarily driven by PR change size and commit features, and Devin by review and discussion features.

%
%
%
Such distinctions are  reflected in the feature-level coefficients of our models as well. For example, within the PR change size and commit family, a unit increase in the number of commits linked to a PR increases merge likelihood by 2.11 times for Copilot and 1.57 times for OpenAI Codex, but is associated with an approximately 7\% decrease in merge odds for Devin. Similarly, a higher volume of submitter-reviewer interaction increases the odds of merging for OpenAI Codex-authored PRs, compared to others.

\begin{framed}
\noindent\textbf{Implication 3.} Project maintainers may select agents based on the specific characteristics of their development workflows. Consequently, researchers should investigate how to design PR review workflows that better leverage these varying agent characteristics.
\end{framed}







\section{Threats to Validity}
PRs with NULL \texttt{merged\_at} values include both closed and open PRs; the latter may introduce noise, though the large sample size mitigates this effect.
Our models identify correlations, not causal relationships.
The \textsc{AIDev} dataset naturally excludes repositories that prohibit AI-generated PRs.
Additionally, Cursor and Claude Code were excluded from RQ3 due to small sample sizes ($N$=1,541 and $N$=459), limiting generalizability of our results for these agents.

\section{Discussion and Future Work}

Our findings reveal both shared and distinct factors influencing the merge outcomes of human-authored and agent-authored PRs, suggesting that uniform PR creation and review workflows may not be optimal for both.
For instance, Khare et al.~\cite{khare2025deputydevaipowered} show that hybrid code review workflows---where AI handles routine checks and humans focus on higher-level judgment---better leverage complementary strengths.
For human-AI collaboration, Hassan et al.~\cite{hassan2024rethinkingsoftwareengineeringfoundation} advocate a shift from task-first approaches to an approach where humans and AI discuss and refine goals through dialogues.
Future work may explore collaborative development that delineates human and AI roles, leveraging AI for PR generation while preserving human contributions in reviews.

\section*{Acknowledgments}
This work was supported by JST SPRING Grant Number JPMJSP2140.

\newpage
\balance
\bibliographystyle{ACM-Reference-Format}
\bibliography{references}

\end{document}